\documentclass[twocolumn,showpacs,aps,epsfig,nofootinbib]{revtex4}

\usepackage{graphicx}
\usepackage{color}
\usepackage{epstopdf}
\usepackage{latexsym}
\usepackage{amssymb}
\usepackage{color}

\usepackage[center]{subfigure}

\begin{document}

 \newcommand{\bq}{\begin{equation}}
 \newcommand{\eq}{\end{equation}}
 \newcommand{\bqn}{\begin{eqnarray}}
 \newcommand{\eqn}{\end{eqnarray}}
 \newcommand{\nb}{\nonumber}
 \newcommand{\lb}{\label}
\newcommand{\PRL}{Phys. Rev. Lett.}
\newcommand{\PL}{Phys. Lett.}
\newcommand{\PR}{Phys. Rev.}
\newcommand{\CQG}{Class. Quantum Grav.}

\title{Holographic Superconductors in a Rotating Spacetime}

\author{Kai Lin}
\email{lk314159@hotmail.com}

\author{E. Abdalla}
\email{eabdalla@usp.br}

\affiliation{Instituto de F\'isica, Universidade de S\~ao Paulo, CP
66318, 05315-970, S\~ao Paulo, Brazil}

\date{\today}

\begin{abstract}
We consider holographic superconductors in a rotating black string spacetime. In view of the mandatory introduction of the  $A_\varphi$ component of the vector potential we are left with three equations to be solved. Their solutions show that the effect of the rotating parameter $a$ influences the critical temperature $T_c$ and the conductivity $\sigma$ in a simple but non trivial way.

\end{abstract}


\maketitle

\section{Introduction}
\renewcommand{\theequation}{1.\arabic{equation}} \setcounter{equation}{0}

The use of the AdS/CFT (Anti de Sitter/Conformal Field Theory)
correspondence has been recently proposed  by  Hartnoll, Herzog and
Horowitz for the investigation of the strongly correlated condensed
matter physics from the gravitational dual\cite{holographic}. It
generally shows that the instability of black strings corresponds to
a second-order phase transition from the normal state to a
superconductor state with the spontaneous $U(1)$ symmetry breaking
and leads to interesting physics in the related lower dimensional
physics.

Such a breakthrough result has been widely used more recently to
model conductivity and other condensed matter physics properties, in
particular, holographic superconductor in various
spacetimes\cite{everybody}. Most authors considered, up to now, the
holographic superconductor on a static background. However, we still
have room for the other hair, namely angular momentum, on the top of
mass and charge. Or, equivalently, we can say that a real black
string (or black hole) could  have angular momentum in the
background. It is thus our goal to study the holographic
superconductor related to the 4 dimensional rotating black hole. We
consider the simplest uncharged background case and we will use both
analytical and numerical methods to find the critical temperature
and conductivity of the holographic superconductor related to the
rotating black hole background.

The paper is planned as follows. In section II, we introduce the
metric of the uncharged rotating black hole and calculate the
Hawking radiation and Hawking temperature. Then, the analytical and
numerical methods are applied to find the conductivity of such a
black hole in sections III and IV respectively. In section V and VI,
we want to improve the analytical method to research the holographic
superconductor with backreactions. Section VII includes a summary
and a conclusion.

\section{Uncharged Rotating black strings and Hawking Temperature}
\renewcommand{\theequation}{2.\arabic{equation}} \setcounter{equation}{0}

The simplest uncharged rotating black string in 4 dimensional ADS spacetime is given
by\cite{lemos}\cite{Metric}
 \bqn
  \lb{metric}
ds^2=-f(r)\left(\Xi dt-a
d\varphi\right)^2+r^2\left(\frac{a}{l^2}dt-\Xi
d\varphi\right)^2&&\nb\\
+\frac{dr^2}{f(r)}+\frac{r^2}{l^2}d\theta^2,~~~~~~~&&
 \eqn
where $f(r)=\frac{r^2}{l^2}-\frac{2M}{r}$ and
$\Xi=1+\frac{a^2}{l^2}$; $M$ and $a$ are the mass and rotation
parameters of the black string, while $\Lambda =-3/l^2$ is the cosmological
constant. Without loss of generality, we set $l=1$ and rewrite the function
$f(r)$ as
 \bqn
  \lb{metricf}
f(r)=r^2-\frac{r_h^3}{r},
 \eqn
where $r_h$ is event horizon of the AdS black hole. Although a very simple model of rotation, the results will show dependence on the above rotation parameter $a$.
Rotation is one of the few possible hairs of a Black Hole and corresponds to a general solution of the end fate of a collapsing star. The solution we are considering is 
well simpler than the full rotating Kerr solution, but is a sensible and consistent model. On the other hand, rotation in a physical condensed matter system 
is generally related to spin and corresponds also to a generic physical system.

We first find the Hawking radiation and temperature
of such a black hole solution. A simple and effective method to
study the Hawking radiation is the tunneling radiation theory of
black holes proposed by Robinson and Wilczek et al \cite{Tunneling}.
Developing the tunneling theory, the Hamilton-Jacobi method has been
proposed to compute the Hawking tunneling rate and Hawking radiation
at the horizon of the black hole \cite{HJ}. We use that method to
find the Hawking tunneling radiation and the temperature of the
black hole.

Let us consider the semi-classical Hamilton-Jacobi equation \cite{HJ}
 \bqn
  \lb{HJ}
g^{\mu\nu}\frac{\partial S}{\partial x^{\mu}}\frac{\partial
S}{\partial x^{\nu}}+\mu_0^2=0,
 \eqn
where $\mu_0$ is the mass of the radiation particle. We shall use a
semi classical approximation to find the Hawking radiation and
temperature of the Black Hole.

In the rotating black string spacetime, we separate the Hamilton-Jacobi equation as
 \bqn
  \lb{S}
S=-\omega_0 t+R(r)+m_0\varphi+Y(\theta),
 \eqn
where, $\omega_0$ and $m_0$ are the energy and angular momentum of particles respectively, so that we obtain the radial and angular Hamilton-Jacobi equations as
 \bqn
  \lb{RHJ}
&&f^2R'^2-\frac{(\Xi\omega_0-m_0a)^2}{(a^2-\Xi^2)^2}\nb\\
&&~~~~~~+f\left[\frac{(a\omega_0-m_0\Xi)^2}{r^2(a^2-\Xi^2)^2}+\frac{\lambda}{r^2}+\mu_0^2\right]=0\quad ,\nb\\
&&\left(\frac{dY}{d\theta}\right)^2=\lambda\quad ,
 \eqn
where $\lambda$ is a constant. At the event horizon $r_0$, we can
expand the function $f$ as
 \bqn
  \lb{Expandf}
f(r)=f'(r_h)(r-r_h)+\frac{f''(r_h)}{2}(r-r_h)^2+\cdot\cdot\cdot\quad ,
 \eqn
and get
 \bqn
  \lb{RHJ2}
R_\pm=\pm\frac{i\pi}{f'(r_h)}\frac{\omega_0-\frac{m_0a}{\Xi}}{\Xi-\frac{a^2}{\Xi}}\quad ,
 \eqn
where $R_+$ and $R_-$ are the radial outgoing and incoming modes,
respectively \cite{Tunneling}. Therefore, the tunneling rate at the event horizon is
 \bqn
  \lb{Tunnelingrate}
\Gamma=e^{-2(\Im R_+-\Im
R_-)}=\exp\left(-\frac{4\pi}{f'(r_h)}\frac{\omega_0-\frac{m_0a}{\Xi}}{\Xi-\frac{a^2}{\Xi}}\right)\quad ,
 \eqn
while the Hawking temperature is given by \cite{Tunneling}
 \bqn
  \lb{Temperature}
T_h=\frac{f'(r_h)}{4\varrho\pi}=\frac{3r_h}{4\varrho\pi}
 \eqn
where $\varrho=\frac{\Xi}{\Xi^2-a^2}=\frac{1+a^2}{1+a^2+a^4}$.

We thus find that the form of the Hawking temperature is very
similar to the temperature of static black hole, while the
$\varrho$ in Eq.(\ref{Temperature}) depends on the rotation
parameter $a$. We use the results to analyse the holographic
superconductor in the next section.

\section{Holographic Superconductor Modes and Analytical Investigation}
\renewcommand{\theequation}{3.\arabic{equation}} \setcounter{equation}{0}
In the rotating black hole spacetime background, the Lagrangian
density of the simplest holographic superconductor model with a
Maxwell field and a charged complex scalar field is
 \bqn
  \lb{action}
{\cal L}=-\frac{1}{4}F^{\mu\nu}F_{\mu\nu}-|\partial
\Psi-iA\Psi|^2-\frac{m^2}{L^2}\Psi^2\quad ,
 \eqn
where $F_{\mu\nu}=\partial_\mu A_\nu-\partial_\nu A_\mu$ and
$\Psi=\Psi(r)$ since $g_{\mu\nu}$ only depends on $r$. As in the
static background spacetime case, $A_\mu=\delta^t_\mu \Phi(r)$, but
in the rotating black string background, we must set
$A_\mu=\delta^t_\mu \Phi(r)+\delta^\varphi_\mu \Omega(r)$ in view of
the presence of the $g_{t\varphi}$ term.

From the variation of Eq.(\ref{action}), we  get three equations. They are
 \begin{eqnarray}
\Psi''+\left(\frac{f'}{f}+\frac{2}{r}\right)\Psi'+\left[\frac{(1+a^2)\Phi+a\Omega}{(1+a^2+a^4)f}\right]^2\Psi&&\nb\\
+\left[\frac{(1+a^2)\Omega+a\Phi}{r(1+a^2+a^4)f}\right]^2\Psi-\frac{m^2}{L^2f}\Psi&=&0~~~~~~\lb{FEQ1}\\
\Phi''+\left[\frac{2(1+a^2)^2}{(1+a^2+a^4)r}-\frac{a^2f'}{(1+a^2+a^4)f}\right]\Phi'~~~&&\nb\\
-\frac{2\Psi^2}{f}\Phi+\left[\frac{2(a+a^3)}{(1+a^2+a^4)r}-\frac{a(1+a^2)f'}{(1+a^2+a^4)f}\right]\Omega'&=&0  \lb{FEQ2}\\
\Omega''-\left[\frac{2a^2}{(1+a^2+a^4)r}-\frac{\left(1+a^2\right)^2f'}{(1+a^2+a^4)f}\right]\Omega'~~~&&\nb\\
-\frac{2\Psi^2}{f}\Omega-\left[\frac{2(a+a^3)}{(1+a^2+a^4)r}-\frac{a(1+a^2)f'}{(1+a^2+a^4)f}\right]\Phi'&=&0  \lb{FEQ3}
 \end{eqnarray}
The boundary condition at the horizon requests $A_\mu$ to be finite.
The solutions of above equations at infinity are given by
 \bqn
  \lb{boundary}
\Psi&=&\frac{\sqrt{2} \langle{\cal O}_1\rangle }{r^{\Delta_-}}+\frac{\sqrt{2} \langle{\cal O}_2\rangle }{r^{\Delta_+}}+\cdot\cdot\cdot\quad ,\nb\\
\Phi&=&\mu-\frac{\rho}{r}+\cdot\cdot\cdot\quad ,\nb\\
\Omega&=&\nu-\frac{\zeta}{r}+\cdot\cdot\cdot\quad ,
 \eqn
where $\Delta_\pm=\frac{3}{2}\pm\sqrt{\frac{9}{4}+\frac{m^2}{L^2}}$
implying that $\Delta$ satisfies $\frac{1}{2}<\Delta<3$. In the
following, we set $L=1$.

Siopsis and Therrien proposed an effective analytic method to
calculate the critical temperature of holographic superconductor in
static spacetime\cite{Analytic}. We generalize such a method for the
rotating black hole case in this section.

According to that procedure, we change the coordinates as $z=r_0/r$ and the field equations can be rewritten as
 \bqn
  \lb{changeq1}
z\Psi_{,zz}-\frac{2+z^3}{1-z^3}\Psi_{,z}+\frac{z \left[\left(a^2+1\right) \Phi +a \Omega\right]^2}{\left(a^4+a^2+1\right)^2 r_0^2 \left(1-z^3\right)^2}\Psi \nb\\
   +\frac{z \left(\left(a^2+1\right) \Omega +a\Phi \right)^2}{\left(a^4+a^2+1\right)^2 r_0^2\left(z^3-1\right)}\Psi -\frac{m^2}{z\left(1-z^3\right)}\Psi =0, \eqn
    \bqn
  \lb{changeq2}
\Phi_{,zz}-\frac{3 a^2 z^2 \Phi_{,z}}{\left(a^4+a^2+1\right)
   \left(z^3-1\right)}&-&\frac{2 \Psi^2}{ z^2\left(1-z^3\right)}\Phi\nb\\
   -\frac{3 a \left(a^2+1\right) z^2 \Omega_{,z}}{\left(a^4+a^2+1\right) \left(z^3-1\right)}&=&0 \quad,
      \eqn
    \bqn
  \lb{changeq3}
  \Omega_{,zz}+\frac{3 \left(a^2+1\right)^2 z^2 \Omega_{,z}}{\left(a^4+a^2+1\right) \left(z^3-1\right)}&-&\frac{2 \Psi^2
   }{z^2\left(1-z^3\right) }\Omega\nb\\
+\frac{3 a \left(a^2+1\right) z^2 \Phi_{,z}}{\left(a^4+a^2+1\right)
   \left(z^3-1\right)}&=&0\quad.
      \eqn

At the critical temperature $T_c$, we have $\Psi=0$, thus Eqs.(\ref{changeq2}) and (\ref{changeq3}) become
 \bqn
  \lb{bana}
  \Phi_{,zz}-\frac{3 a^2 z^2 \Phi_{,z}}{\left(a^4+a^2+1\right)
   \left(z^3-1\right)}&&\nb\\
   -\frac{3 a \left(a^2+1\right) z^2 \Omega_{,z}}{\left(a^4+a^2+1\right) \left(z^3-1\right)}&=&0\nb\\
  \Omega_{,zz}+\frac{3 \left(a^2+1\right)^2 z^2 \Omega_{,z}}{\left(a^4+a^2+1\right) \left(z^3-1\right)}&&\nb\\
+\frac{3 a \left(a^2+1\right) z^2 \Phi_{,z}}{\left(a^4+a^2+1\right)
   \left(z^3-1\right)}&=&0.
 \eqn
Therefore, we can rewrite above equations as
$\Omega_{,z}=\Omega_{,z}\left(\Phi_{,zz},\Phi_{,z}\right)$ and
$\Phi_{,z}=\Phi_{,z}\left(\Omega_{,zz},\Omega_{,z}\right)$, and then
substitute into Eq.(\ref{bana}) again, so that the decoupling
equations are given by
 \bqn
  \lb{decoupling}
\Phi_{,zzz}+\frac{2}{z}\frac{2z^3+1}{z^3-1}\Phi_{,zz}&=&0\nb\\
\Omega_{,zzz}+\frac{2}{z}\frac{2z^3+1}{z^3-1}\Omega_{,zz}&=&0
 \eqn
and the solutions are
 \bqn
  \lb{banasolu}
\Phi&=&\mu-\rho z+C_1 \Big[\sqrt{12}\arctan\left(\frac{1+2z}{\sqrt{3}}\right)\nb\\
&&~~~~~~~~+\ln\left(\frac{1+z+z^2}{(1-z)^2}\right)\Big],\nb\\
\Omega&=&\nu-\zeta z+C_2 \Big[\sqrt{12}\arctan\left(\frac{1+2z}{\sqrt{3}}\right)\nb\\
&&~~~~~~~~+\ln\left(\frac{1+z+z^2}{(1-z)^2}\right)\Big],
 \eqn
Considering the boundary condition at the horizon, we require $\Phi|_{z=1}=\Omega|_{z=1}=0$, so that we may set
 \bqn
  \lb{banaform}
\Phi&=&\lambda r_{hc}(1-z),\nb\\
\Omega&=&\bar{\lambda} r_{hc}(1-z),
 \eqn
where  $\lambda=\frac{\rho}{r_{hc}^2}$ \cite{Analytic}, and $r_{hc}$ is the radius of the horizon at critical temperature. Next, substituting Eq.(\ref{banaform}) into Eq.(\ref{bana}), we find $\bar{\lambda}=-\frac{a}{1+a^2}\lambda$.

According to the idea of Siopsis and Terrien, we introduce
$\Psi(z)=\frac{\langle{\cal
O}_\Delta\rangle}{\sqrt{2}r_h^\Delta}z^\Delta F(z)$ (where
$F|_{z=0}=1$, and $F_{,z}|_{z=0}=0$) to match the boundary
condition, and substitute Eq.(\ref{banaform}) into
Eq.(\ref{changeq1}), which becomes
 \bqn
  \lb{Feq}
-F_{,zz}+\frac{1}{z}\left(\frac{2+z^3}{1-z^3}-2\Delta\right)F_{,z}+\frac{\Delta^2z}{1-z^3}F\nb\\
=\frac{\tilde{\lambda}^2}{(1+z+z^2)^2}F,
 \eqn
 where $\tilde{\lambda}=\frac{\lambda}{1+a^2}$. We observe that the form of Eq.(\ref{Feq}) is the same as the results of \cite{Analytic}, so we can directly use the mathematical conclusions about the eigenvalue $\tilde{\lambda}$
 \bqn
  \lb{sollambda}
\tilde{\lambda}^2=\frac{\int^1_0dz\left\{z^{2\Delta-2}\left[(1-z^3)
\left(F_{,z}\right)^2+\Delta^2zF^2\right]\right\}}{\int^1_0dz\left[z^{2\Delta-2}\frac{1-z}{1+z+z^2}F^2\right]}\quad ,~~~
 \eqn
and assume a very simple trial function $F=F_\alpha(z)\equiv1-\alpha z^2$. We can thus compute the minimizing value of $\tilde{\lambda}^2$. Then, using the temperature's formula, the critical temperature is given by the expression
  \bqn
  \lb{ctemp}
T_c=\frac{3}{4\varrho\pi}r_{hc}=\frac{3}{4\varrho\pi}\sqrt{\frac{\rho}{\lambda}}=\frac{3}{4\pi}\eta\sqrt{\frac{\rho}{\tilde{\lambda}}},
 \eqn
where
  \bqn
  \lb{eta}
  \eta=\frac{1}{\varrho\sqrt{1+a^2}}=\frac{1+a^2+a^4}{(1+a^2)^{3/2}}.
  \eqn
From above equations, we can get the  relation between critical temperature $T_c$ and $\tilde{\lambda}$ which depends on $\Delta$.

In table I we write the results of the analytical method in the rotating black strings spacetime.

\begin{table}[!h]
\tabcolsep 0pt \vspace*{-12pt}
\begin{center}
\def\temptablewidth{0.25\textwidth}
{\rule{\temptablewidth}{1pt}}
\begin{tabular*}{\temptablewidth}{@{\extracolsep{\fill}}ccc}
      $~~~~~\Delta$&~~~~~~&$T_c~~~~~$ \\
\hline
      $~~~~~0.6$&~&$0.45504~\eta\sqrt{\rho}~~~~~$\\
      $~~~~~0.8$&~&$0.29124~\eta\sqrt{\rho}~~~~~$\\
      $~~~~~1$&~&$0.22496~\eta\sqrt{\rho}~~~~~$\\
      $~~~~~1.2$&~&$0.18638~\eta\sqrt{\rho}~~~~~$\\
      $~~~~~1.4$&~&$0.16066~\eta\sqrt{\rho}~~~~~$\\
      $~~~~~1.6$&~&$0.14214~\eta\sqrt{\rho}~~~~~$\\
      $~~~~~1.8$&~&$0.12810~\eta\sqrt{\rho}~~~~~$\\
      $~~~~~2$&~&$0.11704~\eta\sqrt{\rho}~~~~~$\\
      $~~~~~2.2$&~&$0.10809~\eta\sqrt{\rho}~~~~~$\\
      $~~~~~2.4$&~&$0.10067~\eta\sqrt{\rho}~~~~~$\\
      $~~~~~2.6$&~&$0.09440~\eta\sqrt{\rho}~~~~~$\\
      $~~~~~2.8$&~&$0.08903~\eta\sqrt{\rho}~~~~~$\\
       \end{tabular*}
       {\rule{\temptablewidth}{1pt}}
       \end{center}
\caption{Relation of the critical Temperature $T_c$ and $\Delta$.}
\end{table}

It is clearly evident that the conclusion is no other than the
results of \cite{Analytic} as $\eta=1$ ($a=0$), and all the
correction depend on $\eta$, thus we focus on $\eta$.

\begin{figure}
\includegraphics[width=4.3cm]{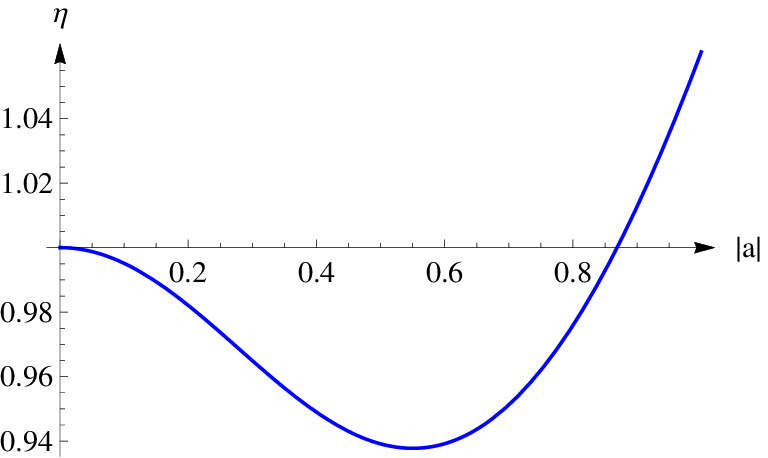}\includegraphics[width=4.3cm]{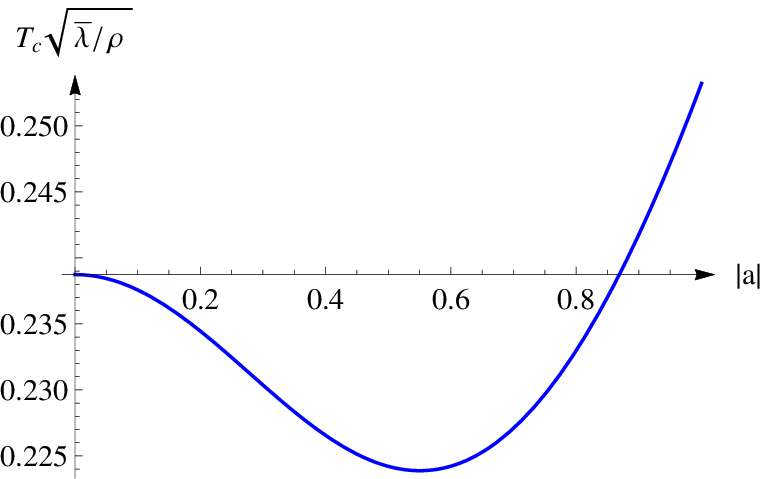}
\caption{The relationship between $\eta$ and $|a|$ according to Eq.(\ref{eta}) and in the second diagram the dependence of the temperature with $\vert a\vert$.}
\label{fig0}
\end{figure}
In Fig. \ref{fig0} we plot the relationship between $\eta$ and $a$. We find that $\eta<1$ as $|a|<a_1$ where $a_1=\sqrt{\frac{2^{2/3}}{6}\left[\left(25+3\sqrt{69}\right)^{1/3}+\left(25-3\sqrt{69}\right)^{1/3}\right]-\frac{1}{3}}\approx 0.868837$. In this region, the correction from $a$ led the critical temperature $T_c$ to decrease. The minimum value of $\eta$ is $\eta_{min}=\sqrt{\frac{26\sqrt{13}}{27}-\frac{70}{27}} \approx 0.937774$ at $a_{min}=\sqrt{\frac{\sqrt{13}-3}{2}}\approx 0.550251$. However, the effect of $\eta$ leads to an increase of the critical temperature $T_c$ as $|a|>a_1$.

Further details of holographic superconductor shall be investigated by a numerical method in the next section.

\section{Numerical Investigation and Conductivity}
\renewcommand{\theequation}{4.\arabic{equation}} \setcounter{equation}{0}

In this section, we set $m^2=-2$, $r_h=1$, and then use the the shooting method \cite{holographic}, combining with the boundary condition (\ref{boundary}) and the main equations (\ref{FEQ1})-(\ref{FEQ3}), to calculate the condensate as a function of temperature. The results are shown in Fig. 2.

\begin{figure*}
\includegraphics[width=8cm]{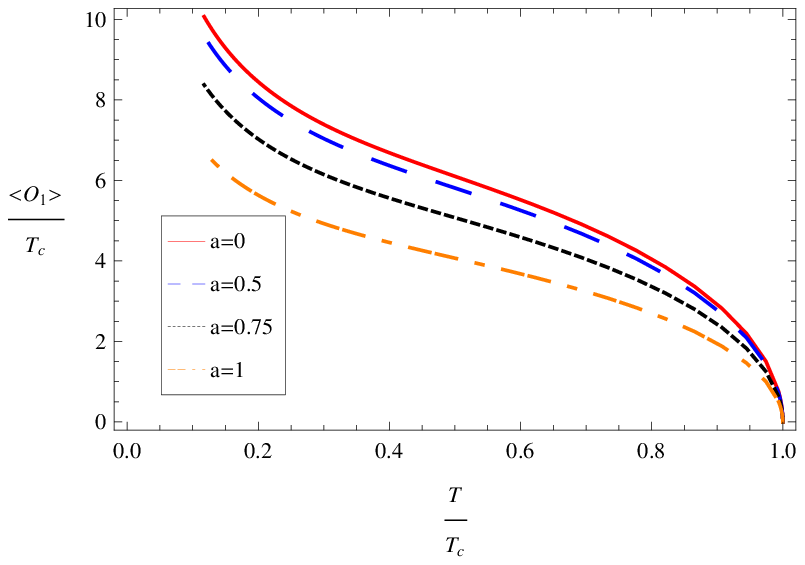}\includegraphics[width=8cm]{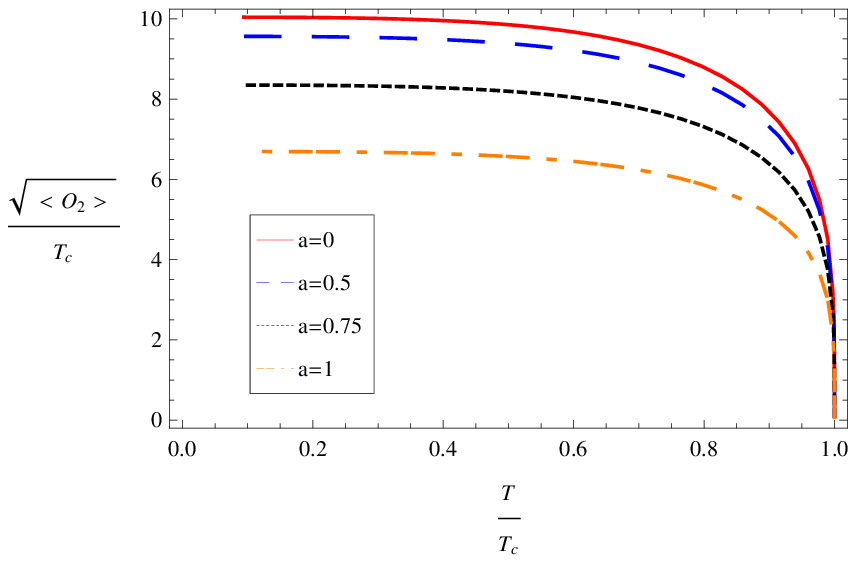}
\caption{The condensate as a function of the temperature for the two
operators ${\cal O}_1$ and ${\cal O}_2$ in rotating spacetime.}
\end{figure*}

In Fig. 2, it is easy to find that, as the rotating parameter $a$ increases, the curved line gets lower.

Finally, we study the conductivity. Considering the perturbed Maxwell field $dA=A_x(r)e^{-i\omega t}dx$, we obtain the equation
 \bqn
  \lb{Axeq}
A_x''+\frac{f'}{f}A_x'-\varrho^2\left[1-\frac{fa^2}{(1+a^2)^2r^2}\right]\frac{\omega^2}{f^2}A_x\nb\\
-2\frac{\Psi^2}{f}A_x=0
 \eqn
The boundary condition at event horizon requires
 \bqn
  \lb{Axbhorizon}
A_x(r)\sim f(r)^{-i\varrho\omega/3r_h},
 \eqn
 while the behavior of $A_x$ in the asymptotic AdS region is
 \bqn
  \lb{AxbADS}
A_x=A^{(0)}+\frac{A^{(1)}}{r}.
 \eqn
 Therefore, we can get the conductivity of the superconductor by using the AdS/CFT dictionary \cite{holographic}
  \bqn
  \lb{AxbADS2}
\sigma=-\frac{iA^{(1)}}{\omega A^{(0)}}.
 \eqn

We use the above equations to calculate the conductivity in Figs. 3 and
4. It can be seen that the effect of $a$ drives  the real  part of the conductivity smaller and smaller. For the imaginary part the behaviour is more involved: the frequency times the imaginary part increases in absolute value.

\begin{figure*}
\includegraphics[width=8cm]{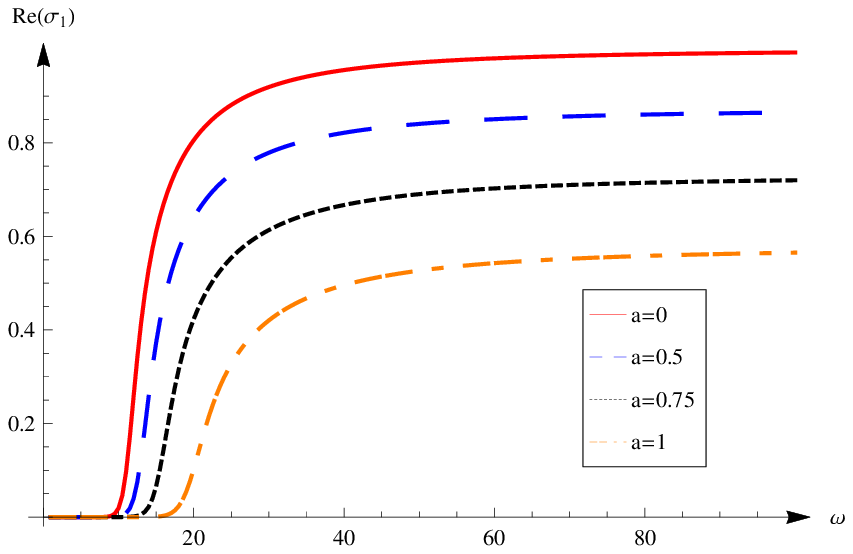}\includegraphics[width=8cm]{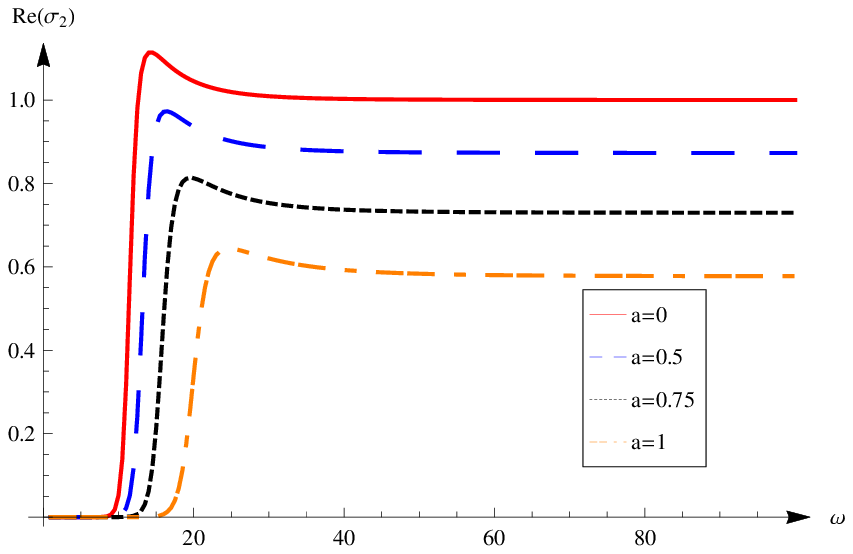}
\caption{The real part of conductivity
for the two operators ${\cal O}_1$ and ${\cal O}_2$ in rotating
spacetime.}
\end{figure*}

\begin{figure*}
\includegraphics[width=8cm]{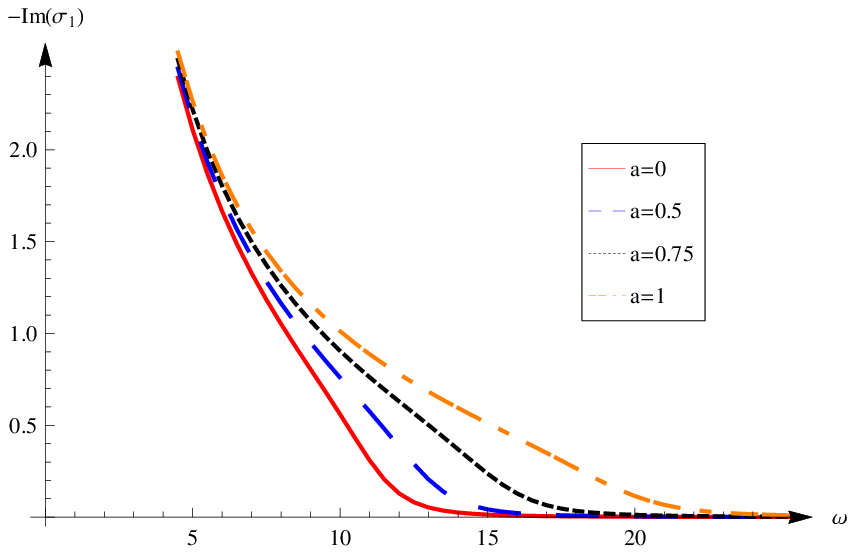}\includegraphics[width=8cm]{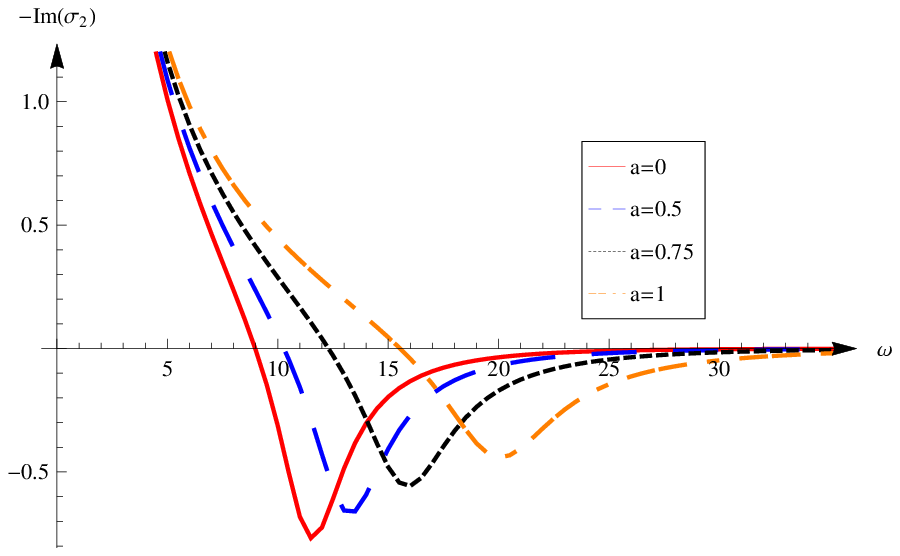}
\caption{The imaginary part of conductivity for the two operators ${\cal O}_1$ and ${\cal O}_2$ in
rotating spacetime.}
\end{figure*}

We also plot $-\omega \text{Im} \sigma$ with small $\omega$, and the results show that $-\omega \text{Im} \sigma$ with different $a$ go to the same constant as $\omega$ goes to $0$. This implies superconductivity for zero frequency.

\begin{figure*}
\includegraphics[width=8cm]{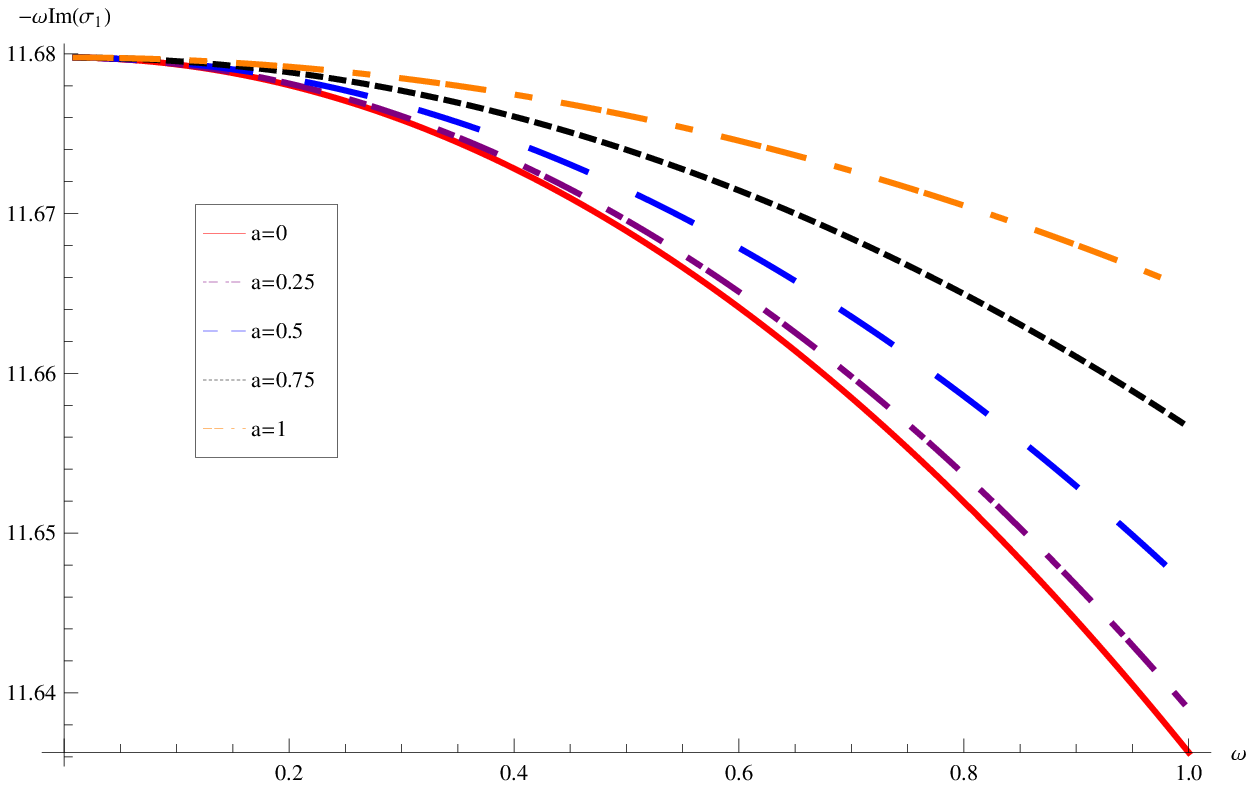}\includegraphics[width=8cm]{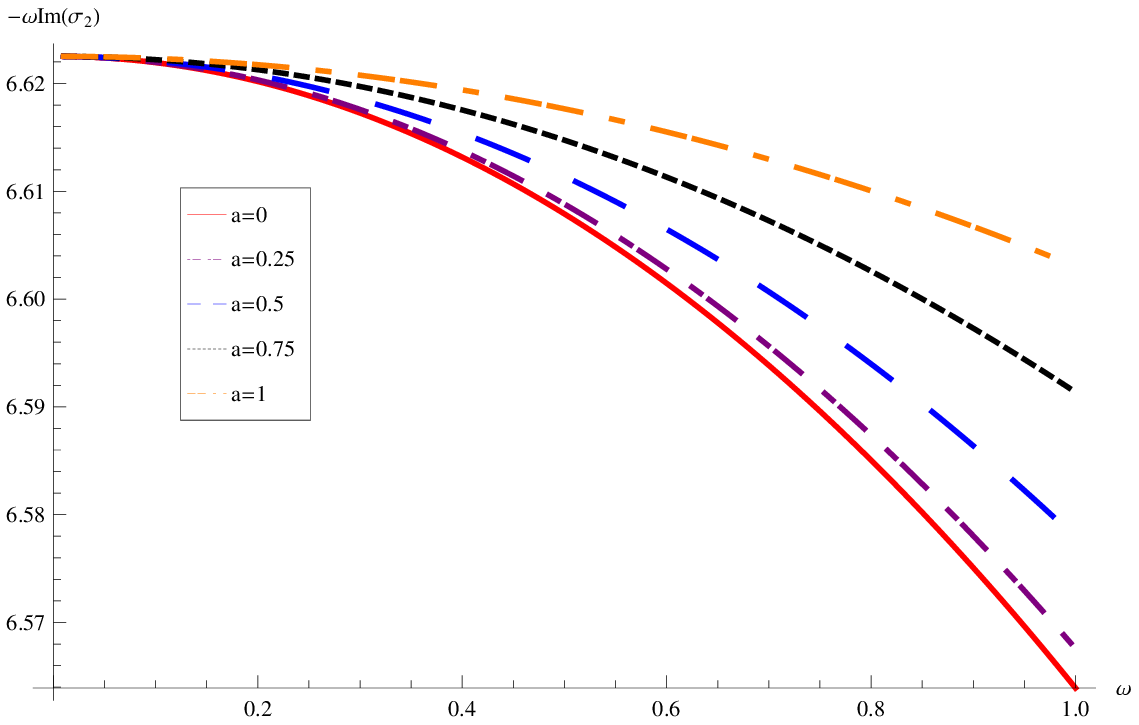}
\caption{$-\omega \text{Im}\sigma$ for the two operators ${\cal O}_1$ and ${\cal O}_2$ in
rotating spacetime.}
\end{figure*}

\section{Modified analytical method with backreaction}
\renewcommand{\theequation}{5.\arabic{equation}} \setcounter{equation}{0}

The authors of reference \cite{backPan} proposed a method to analytically research
the holographic superconductor with backreactions. However, using such a
method still poses a problem: the error will be larger
as the backreaction increases. Here, we improve the
method in \cite{backPan} to get more precise results and use the
improved method to study the holographic superconductors with
backreactions in a rotating black strings spacetime.

Let us consider the general action of holographic superconductor
model in the $d$-dimensional spacetime,
 \bqn
  \lb{action2}
S=\int
d^dx\sqrt{-g}\bigg[\frac{1}{2\kappa^2}\left(R-2\Lambda\right)-\frac{1}{4}F^{\mu\nu}F_{\mu\nu}&&\nb\\
-\left|\partial \psi-iA\psi\right|^2-m^2\psi^2\big],&&
 \eqn
where $\kappa^2=8\pi G_d$, $\Lambda=-\frac{1}{2L^2}(d-1)(d-2)$,
$dA=\phi(r)dt$ and $\psi(r)$ are the $d$-dimensional gravitational
constant, the cosmological constant, the gauge field and the scalar
field respectively.

In this section, we use $d$-dimensional static spacetime as an
example to show the improved analytical method of holographic
superconductor. In \cite{backPan}, authors focus on the
$d$-dimensional planar Anti-de Sitter black hole
 \bqn
  \lb{metric0}
ds^2=-f(r)e^{-\chi(r)}dt^2+\frac{dr^2}{f(r)}+r^2h_{ij}dx^idx^j,
 \eqn
where $h_{ij}dx^idx^j$ is the line element of $(d-2)$-dimensional
hypersurface with the curvature $k=0$.

Considering the backreaction from spacetime, we fail to directly get
the results of $f(r)$ and $\chi(r)$, while the effect from $\psi$ is
very small, so that we can expand them as
 \bqn
  \lb{expand0}
f&=&f_0+\epsilon^2f_2+\epsilon^4f_4+\cdot\cdot\cdot\nb\\
\chi&=&\chi_0+\epsilon^2\chi_2+\epsilon^4\chi_4+\cdot\cdot\cdot
 \eqn
where $\epsilon\ll 1$. According to the calculation in
\cite{backPan}, at the critical temperature,
 \bqn
  \lb{phi0}
\phi_c(z)=\lambda r_c(1-z^{d-3}),
 \eqn
where $\phi_c$ and $r_c$ are $A_t=\phi$ and event horizon position
at critical temperature, while $z=\frac{r_c}{r}$ and
$\lambda=\rho/r_c^{d-2}$ ($\rho$ is the charge density). Therefore,
we can get $f_0$ and $\chi_0$,
 \bqn
  \lb{f0chi0}
f_0&=&r_c^2g(z)\nb\\
&=&r_c^2\left[\frac{1}{L^2z^2}-\frac{z^{d-3}}{L^2}-\kappa^2\lambda^2\frac{d-3}{d-2}z^{d-3}\left(1-z^{d-3}\right)\right]\; ,\nb\\
\chi_0&=&0\quad .
 \eqn
At first order, we can set $\psi=\frac{\psi_C}{r^\Delta}z^\Delta
F(z)$ (where $m^2=\Delta(1-d+\Delta)$ and $\psi_0$ is a constant)
and the asymptotic Anti-de Sitter boundary condition for $\psi$
requests $F(0)=1$ and $F'(0)=0$. Finally, the equation of motion for
$F(z)$ in \cite{backPan} is given by
 \bqn
  \lb{F0}
F''+\left[\frac{2(\Delta+2)-d}{z}+\frac{g'}{g}\right]F'+\Bigg[\lambda^2\frac{(1-z^{d-3})^2}{z^4g^2}\nb\\
+\frac{\Delta}{z}\left(\frac{\Delta+3-d}{z}+\frac{g'}{g}\right)-\frac{m^2}{z^4g}\Bigg]F=0.
 \eqn
From the Sturm-Liouville eigenvalue problem, we get
 \bqn
  \lb{lambda0}
\lambda^2=\frac{\int_0^1T(F'^2-UF^2)dz}{\int_0^1TVF^2dz},
 \eqn
where
 \bqn
  \lb{TUV0}
T&=&z^{2(\Delta+2)-d}g,\nb\\
U&=&\frac{\Delta}{z}\left(\frac{\Delta+3-d}{z}+\frac{g'}{g}\right)-\frac{m^2}{z^4g},\nb\\
V&=&\frac{(1-z^{d-3})^2}{z^4g^2}.
 \eqn
and we assume $F(z)=1-az^2$, which satisfies the AdS boundary
condition.

At the probe limit, we can calculate $\lambda^2$ at critical temperature
as the minimum eigenvalue in Eq.(\ref{TUV0}), and compute the
critical temperature $T_c$ by the relationship between $\lambda$ and
$T_c$.

However, when considering the backreaction from spacetime, we must
face the troubling fact that $g$ in Eq.(\ref{f0chi0}) include $\lambda^2$
as $\kappa\not=0$. In order to solve this difficulty, authors in
\cite{backPan} express $\kappa$ as
 \bqn
  \lb{kappa0}
\kappa_n=n\Delta\kappa,~~~~~n=0,1,2\cdot\cdot\cdot,
 \eqn
where $\Delta\kappa=\kappa_{n+1}-\kappa_n$ and $\kappa_0=0$, so we
can use the $\lambda^2|_{\kappa_{n-1}}$ to replace
$\lambda^2$ in $g(r)$, which can be rewritten as
 \bqn
  \lb{g0}
g\approx
\frac{1}{L^2z^2}-\frac{z^{d-3}}{L^2}-\kappa_n^2\left(\lambda^2|_{\kappa_{n-1}}\right)\frac{d-3}{d-2}z^{d-3}\left(1-z^{d-3}\right)\;.\nb\\
 \eqn
The authors of \cite{backPan} got good results by using above method, but the error
between exact numerical values and the values from above method are
larger as $\kappa$ increases.

We want to use iteration method to improve above method, so let's
consider the $\lambda^2|_{\kappa_{n-1}}$ (or directly
$\lambda^2|_{\kappa_0}$) as the initial value of $\lambda_0^2$ in
$g$ and then calculate the minimum eigenvalue of $\lambda_1^2$ in
Eq.(\ref{lambda0}), and substitute $\lambda_1^2$ into
Eq.(\ref{lambda0}) again to calculate the minimum eigenvalue
$\lambda_2^2$. Using this iteration process, we can get
$\lambda^2$ as $\lambda_{i+1}^2\approx\lambda_i^2$. Next, from the
relation between $T_c$ and $\lambda^2$ we find
 \bqn
  \lb{g02}
T_c=\frac{1}{4\pi}\left[(d-1)-\kappa^2\lambda^2\frac{(d-3)^2}{d-2}\right]\left(\frac{\rho}{\lambda}\right)^{\frac{1}{d-2}}\; .
 \eqn

\begin{table}[ht]
\caption{\label{TableII} Comparing the $T_c$ results of iteration
method and method of \cite{backPan} with $\Delta\kappa=0.025$ in
$5$-dimensional AdS black hole background, where $m^{2}L^2=-3$}
\begin{tabular}{c c c c}
         \hline
~~$\kappa$~~&~~Iteration Method~~&~~Method of
\cite{backPan}~~&~~Numerical~~
        \\
        \hline
$0$&$0.1962\rho^{1/3}$&$0.1962\rho^{1/3}$&$0.1980\rho^{1/3}$
          \\
$0.05$&$0.1934\rho^{1/3}$&$0.1934\rho^{1/3}$&$0.1953\rho^{1/3}$
          \\
$0.10$&$0.1854\rho^{1/3}$&$0.1853\rho^{1/3}$&$0.1874\rho^{1/3}$
          \\
$0.15$&$0.1726\rho^{1/3}$&$0.1722\rho^{1/3}$&$0.1748\rho^{1/3}$
          \\
$0.20$&$0.1558\rho^{1/3}$&$0.1549\rho^{1/3}$&$0.1580\rho^{1/3}$
          \\
$0.25$&$0.1361\rho^{1/3}$&$0.1345\rho^{1/3}$&$0.1382\rho^{1/3}$
          \\
$0.30$&$0.1147\rho^{1/3}$&$0.1123\rho^{1/3}$&$0.1165\rho^{1/3}$
          \\
        \hline
\end{tabular}
\end{table}

In Table II, the numerical results is exact
\cite{backPan}\cite{backH}. we compare the results of iteration
method and the results in \cite{backPan}, and find the values from
iteration method is better.

In section VI, we will use the iteration method to analytically
research the holographic superconductor in a rotating black string
spacetime.

\section{Analytical investigation of holographic superconductors with backreactions}
\renewcommand{\theequation}{6.\arabic{equation}} \setcounter{equation}{0}
The metric and electromagnetical potential of rotating ADS black
strings spacetime could be given by
 \bqn
  \lb{metric1}
ds^2&=&-f(r)\left[N(r)dt-\frac{\omega}{\alpha^2}d\varphi\right]^2+\frac{dr^2}{f(r)}\nb\\
&&+r^2\left[N(r)d\varphi-\omega dt\right]^2+\alpha^2r^2dz^2,\nb\\
dA&=&N(r)\phi(r)dt-\frac{\omega}{\alpha^2}\phi(r)d\varphi,
 \eqn
where $\omega$ is a constant which is determined by the rotating
parameter, and $\Lambda=-3/\alpha^2$ is the cosmological constant.
Without loss of generality, we set $\alpha=1$.
According to the calculation method in section II, the Hawking
temperature of black hole (\ref{metric1}) is\cite{HJ}
 \bqn
  \lb{TemperatureA}
T_h=\frac{f'(r_h)}{4\hat\varrho\pi},
 \eqn
where $\hat\varrho=\frac{N}{N^2-\omega^2}$. Now, let us substitute the
metric and electromagnetical potential in Eq.(\ref{metric1}) into
the holographic superconductor action (\ref{action2}), and from the
variation of the action (\ref{action2}), we get four field equations
 \bqn
  \lb{Fieldequationfr}
&&\kappa ^2 \Bigg[\frac{2 \omega ^2 \phi ^2 f'  N'}{r^2 \omega ^2 N-r^2 N^3}+\frac{2 \phi (r)^2 N'' \left(r^2 N^2-\omega^2f \right)}{r^2 \left(N^3-\omega^2 N\right)}\nb\\
&&+N' \left(\frac{4 \phi \phi ' \left(r^2 N^2-\omega ^2 f\right)}{r^2\left(N^3-\omega ^2 N\right)}-\frac{4 N\phi^2}{r \omega ^2-r N^2}\right)\nb\\
&&+\frac{2 \omega ^2 \left(f -r^2\right) \phi^2N'^2}{r^2 \left(\omega ^2-N^2\right)^2}+4 f \psi '^2-\frac{4 \psi^2 \phi^2}{f}+4 m^2 \psi^2\nb\\
&&+\phi \left(2 \phi''+\frac{4 \phi'}{r}\right)\Bigg]+f''+\frac{4}{r}f'+\frac{2 f}{r^2}-12\nb\\
&&+\frac{N'' \left(-2 r^2 f \left(\omega ^2-2 N^2\right)+\omega ^2 f^2+r^4 \omega^2\right)}{r^2 \left(N^3-\omega ^2 N\right)}\nb\\
&&-\frac{\omega ^2 \left(f+r^2\right)^2 N'^2}{r^2 \left(\omega ^2-N^2\right)^2}-\frac{N'\left(8 f N^2-4 \omega ^2 f +4 r^2 \omega ^2\right)}{r \omega ^2 N-r N^3}\nb\\
&&+\frac{N'\left[2 \omega ^2 f -2 r^2 \left(\omega ^2-2
N^2\right)\right]}{r^2\left(N^3-\omega ^2 N\right)}f'=0,
 \eqn
  \bqn
  \lb{FieldequationNr}
&&N''+\frac{N'^2 \left(\omega ^2 f+r^2 N^2-\kappa ^2 \omega ^2 \phi^2\right)}{r^2 \omega ^2 N-r^2 N^3}\nb\\
&&+\frac{2 \kappa^2\left(N^2-\omega ^2\right) \left(f^2
\psi'^2+\psi^2 \phi^2\right)}{f^2 N}=0\quad ,
 \eqn
  \bqn
  \lb{Fieldequationphir}
&&\phi''+\phi'\left(\frac{2 NN'}{N^2-\omega^2}+\frac{2}{r}\right)-\frac{\omega^2 \left(r^2-f\right) N'^2}{r^2\left(\omega ^2-N^2\right)^2}\phi\nb\\
&&+\phi\left(\frac{NN''}{N^2-\omega ^2}-\frac{2 NN'}{r
\omega^2-rN^2}-\frac{2 \psi^2}{f}\right)=0\quad ,
 \eqn
  \bqn
  \lb{Fieldequationpsir}
  &&\psi ''(r)+\frac{\psi \left(\phi^2-m^2f\right)}{f^2}\nb\\
  &&+\psi ' \left(\frac{f' }{f}+\frac{2NN' }{N^2-\omega^2}+\frac{2}{r}\right)=0\quad .
 \eqn
The field $\phi$ must remain finite at event horizon, while as
$r\rightarrow\infty$, the boundary condition requires
  \bqn
  \lb{Boundary0}
\psi&=&\frac{\sqrt{2}\langle{\cal O}_-\rangle}{r^{\Delta_-}}+\frac{\sqrt{2}\langle{\cal O}_+\rangle}{r^{\Delta_+}}+\cdot\cdot\cdot,\nb\\
\phi&=&\mu-\frac{\rho}{r}+\cdot\cdot\cdot,\nb\\
f&=&r^2+\cdot\cdot\cdot,\nb\\
N&=&N_0+\cdot\cdot\cdot,
 \eqn
where $\rho$ and $\mu$ are, respectively, the charge density and the
chemical potential in the dual field theory and
$\Delta_\pm=\frac{3}{2}\pm\sqrt{\frac{9}{4}+m^2}$ implying that
$\Delta$ satisfies $\frac{1}{2}<\Delta<3$. Please note that
$N_0=1+\omega^2$ according to the rotating black string solution
given in \cite{Metric}.

An effective analytic method proposed by Siopsis and Therrien can be
used to investigate the holographic superconductor  at the critical
temperature \cite{Analytic}. According to the procedure, we should
change the radial coordinates as $z=r_h/r$, so that the field
equations can be rewritten as
 \bqn
  \lb{Fieldequationfz}
&&\kappa ^2 \Bigg\{\frac{2 \omega ^2 z^2 \phi^2 f_{,z}N_{,z}}{\omega^2 r_h^2 N-r_h^2 N^3}+\frac{2 \phi^2 N_{,zz} \left(r_h^2N^2-\omega^2 z^2 f\right)}{r_h^2 \left(N^3-\omega ^2 N\right)}\nb\\
&&+N_{,z}\left[\frac{4 \phi \phi_{,z} \left(r_h^2 N^2-\omega^2 z^2f\right)}{r_h^2 \left(N^3-\omega^2N\right)}-\frac{4 \omega ^2 zf \phi^2}{r_h^2 \left(N^3-\omega ^2 N\right)}\right]\nb\\
&&-\frac{2\omega ^2 \phi^2 N_{,z}^2 \left(r_h^2-z^2 f\right)}{r_h^2 \left(\omega^2-N^2\right)^2}-\frac{4 r_h^2 \psi^2 \phi^2}{z^4 f}+4 f\psi_{,z}^2\nb\\
&&+\frac{4 m^2 r_h^2 \psi^2}{z^4}+2 \phi\phi_{,zz}\Bigg\}+f_{,zz}\nb\\
&&+f_{,z}\left[\frac{N_{,z} \left(2 r_h^2 \left(\omega ^2-2 N^2\right)-2\omega ^2 z^2 f\right)}{r_h^2 N \left(\omega ^2-N^2\right)}-\frac{2}{z}\right]\nb\\
&&+\frac{z^2 N_{,zz} \left(-\frac{2fr_h^2 \left(\omega ^2-2N^2\right)}{z^2}+\omega^2 f^2+\frac{\omega ^2 r_h^4}{z^4}\right)}{r_h^2 \left(N^3-\omega ^2 N\right)}\nb\\
&&-\frac{\omega^2N_{,z}^2 \left(z^2f+r_h^2\right)^2}{z^2 r_h^2 \left(\omega ^2-N^2\right)^2}-\frac{2 \omega ^2 N_{,z} \left(r_h^4-z^4 f^2\right)}{z^3 r_h^2 \left(N^3-\omega^2 N\right)}\nb\\
&&+\frac{2 \left(z^2 f-6 r_h^2\right)}{z^4}=0\quad ,
 \eqn
 \bqn
  \lb{FieldequationNz}
&&N_{,zz}+\frac{2 N_{,z}}{z}+\frac{N_{,z}^2 \left(\omega^2 z^2 f+r_h^2 N^2-\kappa^2 \omega^2-z^2 \phi^2\right)}{\omega^2 r_h^2 N-r_h^2N^3}\nb\\
&&+\frac{2 \kappa ^2 \left(N^2-\omega^2\right) \left(z^4f^2
\psi_{,z}^2+r_h^2 \psi^2 \phi^2\right)}{z^4 f^2N}=0\quad ,
 \eqn
 \bqn
  \lb{Fieldequationphiz}
&&\phi \left(-\frac{\omega ^2 N_{,z}^2 \left(r_h^2-z^2f\right)}{r_h^2 \left(\omega ^2-N^2\right)^2}-\frac{2 r_h^2 \psi^2}{z^4 f}+\frac{NN_{,zz}}{N^2-\omega^2}\right)\nb\\
&&+\frac{2NN_{,z} \phi_{,z}}{N^2-\omega ^2}+\phi_{,zz}=0\quad ,
 \eqn
 \bqn
  \lb{Fieldequationpsiz}
\psi_{,z}\left(\frac{f_{,z}}{f}+\frac{2 NN_{,z}}{N^2-\omega^2}\right)+\frac{r_h^2 \psi \left(\phi^2-m^2 f\right)}{z^4f^2}+\psi_{,zz}=0\quad .\nb\\
 \eqn
It is very difficulty to calculate the exact solution from above
equations, but considering the fact that the effect from $\psi$ is very
small, at critical temperature we can expand $f$, $N$, $\phi$,
$\psi$ and chemical potential $\mu$ as
 \bqn
  \lb{expand}
\psi&=&\epsilon\psi_1+\epsilon^3\psi_3+\epsilon^5\psi_5+\cdot\cdot\cdot,\nb\\
\phi&=&\phi_0+\epsilon^2\phi_2+\epsilon^4\phi_4+\cdot\cdot\cdot,\nb\\
f&=&f_0+\epsilon^2f_2+\epsilon^4f_4+\cdot\cdot\cdot,\nb\\
N&=&N_0+\epsilon^2N_2+\epsilon^4N_4+\cdot\cdot\cdot,\nb\\
\mu&=&\mu_0+\epsilon^2\mu_2+\epsilon^4\mu_4+\cdot\cdot\cdot,
 \eqn
where $\epsilon=\langle{\cal O}_\pm\rangle\ll1$. Thus, the calculation
accuracy is enough to consider the first terms in Eq.(\ref{expand}).

Near the critical temperature point, the scalar
field $\psi$ vanish and we can use $N_0$ to replace $N$, so that
Eq.(\ref{Fieldequationphiz}) has the solution
 \bqn
  \lb{phiboundary}
A_t\approx N_0\phi_0=\lambda r_{hc}(1-z),
 \eqn
where $\lambda=\rho r_{hc}^{-2}$, and the $r_{hc}$ is the event
horizon position at the critical temperature point. For convenience,
we set $\hat\lambda=\frac{\lambda}{N_0}$ and $\hat
\rho=\frac{\rho}{N_0}$. Next, substituting Eq.(\ref{phiboundary})
into Eq.(\ref{Fieldequationfz}), we obtain
 \bqn
  \lb{fboundary}
f_0=r_{hc}^2\left[z^{-2}-z+\kappa^2\hat\lambda^2\frac{z}{2}(z-1)\right].
 \eqn
Let us set $\psi_1\sim\frac{\langle{\cal
O}_\pm\rangle}{r_{hc}^\Delta}z^{\Delta}F_0(z)$, so that
Eq.(\ref{Fieldequationpsiz}) becomes
 \bqn
  \lb{Fequation}
F_{0,zz}+\left(\frac{f_{0,z}}{f_0}+\frac{2 N_0N_{0,z}}{N_0^2-\omega^2}+\frac{2 \Delta }{z}\right)F_{0,z}&&\nb\\
+\Bigg[\frac{\Delta  f_{0,z}}{zf_0}-\frac{m^2 r_{hc}^2}{z^4 f}+\frac{\hat\lambda^2 (z-1)^2 r_{hc}^4}{z^4 f_0^2}&&\nb\\
-\frac{2 \Delta  N_0N_{0,z}}{\omega ^2 z-zN_0^2}+\frac{(\Delta -1)
\Delta }{z^2}\Bigg]F_0&=&0
 \eqn
Therefore, from the Sturm-Liouville eigenvalue problem, we get, as in (\ref{lambda0}),
 \bqn
  \lb{lambda1}
\lambda^2=\frac{\int_0^1T_0(F_0'^2-U_0F_0^2)dz}{\int_0^1T_0V_0F_0^2dz},
 \eqn
where
 \bqn
  \lb{TUV1}
T_0&=&r_{hc}^2 z^{2 \Delta } \left(\frac{1}{z^2}+\frac{1}{2} \kappa
^2\hat\lambda ^2 (z-1) z-z\right),\nb\\
U_0&=&\frac{4}{\left(\kappa ^2 \hat\lambda ^2 z^3-2 \left(z^2+z+1\right)\right)^2},\nb\\
V_0&=&\frac{\Delta \kappa ^2 \hat\lambda ^2 z^3 (-\Delta +\Delta
z+z)-2 m^2}{\kappa ^2\hat\lambda ^2 z^6-z^5 \left(\kappa ^2 \hat\lambda ^2+2\right)+2 z^2}\nb\\
   &&-\frac{2\Delta \left(\Delta  \left(z^3-1\right)+3\right)}{\kappa ^2\hat\lambda ^2 z^6-z^5 \left(\kappa ^2 \hat\lambda ^2+2\right)+2 z^2},
 \eqn
where we assume $F(z)=1-az^2$, so that we can use the iteration
method to calculate the solution for $\lambda$. Finally, we can
compute the temperature (\ref{TemperatureA}) by
 \bqn
  \lb{TemperatureB}
T_c=\frac{f_0'(r_{hc})}{4\varrho_0\pi}=\frac{6-\kappa^2\hat\lambda^2}{8\varrho_0\pi}r_{hc}=\frac{6-\kappa^2\hat\lambda^2}{8\pi\varrho_0}\sqrt{\frac{\rho}{\hat\lambda}},
 \eqn
where $\varrho_0=\frac{N_0}{N_0^2-\omega^2}$. Comparing with the
critical temperature of statical black strings in \cite{backPan}, we
found all the correction of rotating parameter $\omega$ comes from
$\varrho_0$, while the process of calculating $\hat\lambda$ by
Eq.(\ref{Fequation}) does not depend on $\omega$.

Now, we compare the analytical solution with the exact numerical
solution in Table III. Because the calculation of $\hat\lambda$
does not depend on rotating parameter, while correction of $\omega$
is included in $\varrho_0$, we can set $\varrho_0=1$ without loss of
generality.

\begin{table}[ht]
\caption{\label{TableIII} Comparing the $T_c$ results of iteration
method and Numerical exact solution \cite{backPan}\cite{backH} in
$4$-dimensional AdS black hole background as $m^2=-2$.}
\begin{tabular}{c c c}
         \hline
~~$\kappa$~&~~~~~~~~~~~~~~$\Delta=1$~~~~~~~~~~~~~~&~~~~~~~~~~~~~~$\Delta=2$~~~~~~~~~~~~~~
        \\
\end{tabular}
\begin{tabular}{c c c c c}
         \hline
~~~~&~~Analytical~~&~~Numerical~~&~~Analytical~~&~~Numerical~~
        \\
        \hline
$0$& $0.2250\rho^{1/2}$ & $0.2255\rho^{1/2}$ & $0.1170\rho^{1/2}$ &
$0.1184\rho^{1/2}$
          \\
$0.05$& $0.2249\rho^{1/2}$ & $0.2253\rho^{1/2}$ & $0.1163\rho^{1/2}$
& $0.1177\rho^{1/2}$
          \\
$0.10$& $0.2246\rho^{1/2}$ & $0.2250\rho^{1/2}$ & $0.1142\rho^{1/2}$
& $0.1156\rho^{1/2}$
          \\
$0.15$& $0.2241\rho^{1/2}$ & $0.2245\rho^{1/2}$ & $0.1107\rho^{1/2}$
& $0.1121\rho^{1/2}$
          \\
$0.20$& $0.2235\rho^{1/2}$ & $0.2239\rho^{1/2}$ & $0.1060\rho^{1/2}$
& $0.1074\rho^{1/2}$
          \\
$0.25$& $0.2226\rho^{1/2}$ & $0.2230\rho^{1/2}$ & $0.1003\rho^{1/2}$
& $0.1017\rho^{1/2}$
          \\
$0.30$& $0.2216\rho^{1/2}$ & $0.2220\rho^{1/2}$ & $0.0938\rho^{1/2}$
& $0.0951\rho^{1/2}$
          \\
        \hline
\end{tabular}
\end{table}

On the other hand, we also calculate the $T_c$ with different
$\kappa$ and $\Delta$ in Table IV.

\begin{table}[ht]
\caption{\label{TableIV} Comparing the $T_c$ results of iteration
method and Numerical exact solution in $4$-dimensional AdS black
hole background, where $\hat\eta=\frac{\sqrt\rho}{\varrho_0}$.}
\begin{tabular}{c c c c c}
         \hline
~~~$\Delta$~~~& ~~~$\kappa=0$ ~~~&~~~ $\kappa=0.1$~~~ &
~~~$\kappa=0.2$ ~~~&~~~$\kappa=0.3$~~~
        \\
        \hline
$0.6$& $0.4550\hat\eta$ & $0.4550\hat\eta$ & $0.4549\hat\eta$ &
$0.4547\hat\eta$
          \\
$0.8$& $0.2912\hat\eta$ & $0.2911\hat\eta$ & $0.2906\hat\eta$ &
$0.2898\hat\eta$
          \\
$1$& $0.2250\hat\eta$ & $0.2246\hat\eta$ & $0.2235\hat\eta$ &
$0.2216\hat\eta$
          \\
$1.2$& $0.1864\hat\eta$ & $0.1857\hat\eta$ & $0.1837\hat\eta$ &
$0.1804\hat\eta$
          \\
$1.4$& $0.1607\hat\eta$ & $0.1596\hat\eta$ & $0.1564\hat\eta$ &
$0.1513\hat\eta$
          \\
$1.6$& $0.1421\hat\eta$ & $0.1406\hat\eta$ & $0.1359\hat\eta$ &
$0.1286\hat\eta$
          \\
$1.8$& $0.1281\hat\eta$ & $0.1259\hat\eta$ & $0.1196\hat\eta$ &
$0.1099\hat\eta$
          \\
$2$& $0.1170\hat\eta$ & $0.1142\hat\eta$ & $0.1060\hat\eta$ &
$0.0938\hat\eta$
          \\
$2.2$& $0.1081\hat\eta$ & $0.1045\hat\eta$ & $0.0943\hat\eta$ &
$0.0795\hat\eta$
          \\
$2.4$& $0.1007\hat\eta$ & $0.0962\hat\eta$ & $0.0839\hat\eta$ &
$0.0666\hat\eta$
          \\
$2.6$& $0.0944\hat\eta$ & $0.0890\hat\eta$ & $0.0745\hat\eta$ &
$0.0550\hat\eta$
          \\
$2.8$& $0.0890\hat\eta$ & $0.0827\hat\eta$ & $0.0660\hat\eta$ &
$0.0446\hat\eta$
          \\
        \hline
\end{tabular}
\end{table}

The results show that the critical temperature $T_c$ will decrease
as $\kappa$ or $\Delta$ increase, and from the relationship
$\hat\eta=\frac{\sqrt\rho}{\varrho_0}$, we also find the effect of
rotating parameter $\omega$ also make the $T_c$ decrease.

\section{Conclusions}
\renewcommand{\theequation}{5.\arabic{equation}} \setcounter{equation}{0}

We have considered holographic superconductors in 3+1 dimensional
rotating black strings. The investigation shows that the
$A_\varphi=\Omega$ term  can be ignored in 3+1 dimensional static
spacetime.  The effect from rotating parameter $a$ leads  the real  part of the conductivity to be smaller as the rotation increases. On the other hand, the frequency times the imaginary part increases in absolute value.
Superconductivity, however, remains. This may imply that the presence of spin can eventually 
minimize superconductivity in a physical system.

On the other hand, we proposed an iteration method to improve the
analytical calculation for holographic superconductors with
backreactions, and it is proved that we can use this method to get
better results. We also use this method to research the holographic
superconductor in a rotating spacetime, while the results denote
that the influence of rotating parameter mainly concentrate on
$\varrho_0$ terms, which could make the critical temperature
decrease. At the same time, the backreactions also could lead $T_c$
to decrease.

The spacetime we considered in this paper is rotating black strings
spacetime, while it could be more meaningful to study further rotating black hole case. Some work about
superconducting instability of Kerr-Newmann-de Sitter black hole
appeared in \cite{KNAdS}. Further work on  the holographic superconductor in Kerr-Newmann-de Sitter
spacetime is under way. On the other hand, recently Ho\v{r}ava proposed a new
gravity theory, Ho\v{r}ava-Lifshitz gravity \cite{HL}, which could
solve several difficulty in quantum gravity and cosmology such as
dark energy and dark matter, and our works \cite{PostNewtonian} have
proved this theory satisfy the results of post-Newtonian
approximations, so it will be very interesting to research the
holographic superconductors in Ho\v{r}ava-Lifshitz gravity with
backreactions. Further study is under way.

\section*{\bf Acknowledgements}

This work was supported in part by FAPESP No. 2012/08934-0 and CNPq.

\onecolumngrid

\end{document}